\documentclass[aps,prb,twocolumn,showpacs,achemso,amsmath,amssyb]{revtex4}

\usepackage{graphicx,epsfig}

\begin{document}

\title{Optical pumping of charged excitons in unintentionally doped InAs quantum dots}

\author{G. Mu\~{n}oz-Matutano}

\author{B. Al\'en}
\email[]{benito@imm.cnm.csic.es}
\altaffiliation{Permanent address: IMM, Instituto de Microelectr\'onica de Madrid (CNM, CSIC), Isaac Newton 8,
 28760 Tres Cantos, Madrid, Spain.}

\author{J. Mart\'\i{}nez-Pastor}

\affiliation{ICMUV, Instituto de Ciencia de Materiales, Universidad de Valencia,
  P.O. Box 2085, 46071 Valencia, Spain.}

\author{L. Seravalli}

\author{P. Frigeri}

\author{S. Franchi}

\affiliation{CNR-IMEM Institute, Parco delle Scienze 37/a, 1-43100 Parma, Italy.}

\date{\today}

\begin{abstract}

As an alternative to commonly used electrical methods, we have investigated the optical pumping of charged exciton complexes addressing
impurity related transitions with photons of the appropriate energy. Under these conditions, we demonstrate that the pumping fidelity can
be very high while still maintaining a switching behavior between the different excitonic species. This mechanism has been investigated for
single quantum dots of different size present in the same sample and compared with the direct injection of spectator electrons from nearby
donors.

\end{abstract}


\maketitle

Nowadays, InAs/GaAs self-assembled quantum dots (QDs) are well known nanostructures with important applications envisaged within the
quantum computation and cryptography fields.~\cite{Imamoglu1999, Stevenson2006} The singly charged exciton state (trion), either positive
or negative, is of particular importance because it lacks fine structure splitting, enabling the efficient generation of single photons,
and also because, after radiative recombination, it leaves behind a single charge with well defined spin. Therefore, there is an increasing
interest in the electrical or optical control of the exciton charge state as a necessary step for the spin manipulation.~\cite{Atature2006}
The charge in QD states can be electrically controlled by tuning the gate voltage in field effect structures embedding intrinsic QD
layers.~\cite{Warburton2000,Finley2001,Alen2005} However, this method can produce undesired effects like the reduction of the oscillator
strength induced by the external field.~\cite{Alen2007} The charge state can also be controlled by optical injection, and different
charging schemes have been proposed using above or below barrier
excitation.~\cite{Hartmann2000,Regelman2001,Moskalenko2001,Chang2005,Cade2005} In this work, we demonstrate the selective formation of
charged exciton complexes in initially empty QDs under the presence of unintentional acceptor and donor impurities. Furthermore, the
optical pumping mechanism is investigated for two ensembles of InAs QDs with very different size present in the same sample: small QDs
emitting below 970 nm and large QDs emitting at 1165 nm at 4 K.

\begin{figure}[t]
\includegraphics[width=75 mm]{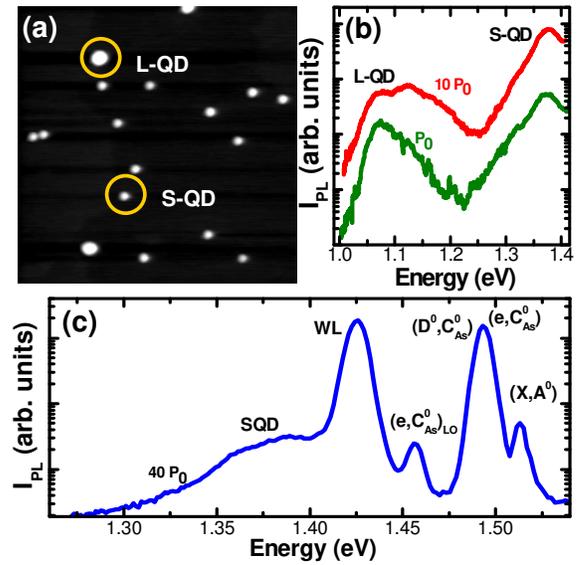}%
\caption{(a) 1x1  $\mu m^2$ AFM image of a similar uncapped sample. Representative QDs of the two families coexisting in the sample have
been encircled; (b) PL spectra obtained at 10 K and two excitation powers showing the emission bands corresponding to each family; (c)
Detail of the PL spectrum in the energy range between the WL and the GaAs edges.} \label{Fig1}
\end{figure}

The MBE (molecular beam epitaxy) growth starts with a 100 nm-thick GaAs buffer grown at 600 ºC, followed by InAs deposition at 505 ºC and
at very low growth rate (0.009 ML/s) and ends with a 100 nm-thick GaAs cap grown by atomic layer MBE at 360 °C.~\cite{Briones1991}During
the InAs deposition, the substrate was not azimuthally rotated, thus producing a continuous variation of InAs coverages on the sample
surface.~\cite{Colocci1997} The combination of low growth rate (LGR) and graded coverage allowed us to obtain particularly low surface
density values, down to 2 $\mu$m$^{-2}$, suitable for optical investigation of isolated QDs. In particular, the coverage of the sample
under consideration here is 2.5 MLs, with a density of about 16 $\mu$m$^{-2}$, as estimated by atomic force microscopy (AFM) measurements
shown in Fig.~\ref{Fig1}(a) carried on uncapped samples.  The AFM images also evidence a bimodal distribution of QD sizes, with most
frequent values of 9 and 14 nm for the heights and 36 and 54 nm for the diameters of small (SQDs) and large (LQDs) quantum dots,
respectively. Both the bimodal size distribution and the relatively large values for the QD dimensions have been reported for similar
nanostructures grown by LGR.~\cite{Costantini2003,Nakata2000}

Conventional photoluminescence (PL) characterization was carried out with the sample held in the cold finger of a closed-cycle He cryostat.
Single QD spectroscopy was performed by using an optical fiber based diffraction limited confocal arrangement inserted in the He exchange
gas chamber of an immersion cryostat. The  PL signal, excited by a tunable Ti:sapphire, was dispersed by a 0.5 (0.3) m focal length grating
spectrograph and detected with a cooled InGaAs focal plane array (Si CCD) for wavelengths above (below) 1000 nm. The excitation of the  PL
spectrum ( PLE) is acquired using the same detectors while varying the excitation wavelength.

Figure~\ref{Fig1}(b) shows the ensemble PL spectra recorded at 10 K using two different excitation power densities (P0 = 0.5 W/cm2) at 790
nm. Two relatively broad emission bands are observed at ~1.08 eV and ~1.38 eV, corresponding to the two different QD families observed by
AFM. Excitation above the GaAs band edge allows also for the observation of the WL line at 1.425 eV and three other bulk related optical
transitions, as shown in Fig.~\ref{Fig1}(c). Three bands are clearly observed corresponding to the GaAs exciton bound to neutral acceptor
at 1.513 eV (X, $C^0$), the free electron-neutral acceptor transition at 1.493 eV (e, $C^0_{As}$), and its LO phonon replica at 1.457
eV.~\cite{Pavesi1994} The unintentional incorporation of impurities (such as carbon acceptors) coming from the growth environment is a
general feature of MBE, as well as of all growth techniques. In our case, Hall measurements of similarly grown GaAs buffer layers reveal a
residual n-type carrier concentration $n=N_D-N_A~10^{15}$ cm$^{-3}$. Thus, the band centered at 1.493 eV is related to the (e, $C^0_{As}$)
and ($D^0$, $C^0_{As}$) recombination paths, and, in thermal equilibrium, a large number of ionized acceptors are available due to the
compensation process enabling the efficient optical pumping of free electrons (and bound holes) as explained below.

\begin{figure}[t]
\includegraphics[width=75 mm]{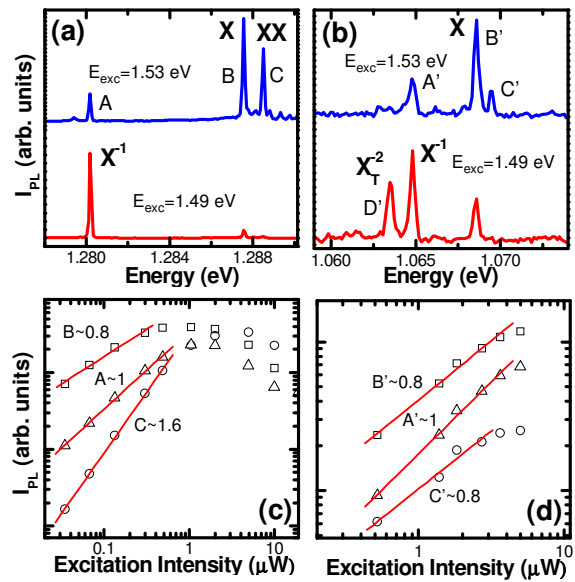}%
\caption{The upper panels show typical  PL spectra obtained at 5 K using two different excitation energies for individual dots of the SQD
(a) and LQD (b) families. Figures (c) and (d) represent the excitation power dependence of the most important transitions under above
barrier excitation for SQD and LQD, respectively.} \label{Fig2}
\end{figure}

Figures~\ref{Fig2}(a) and~\ref{Fig2}(b) show the  PL spectra recorded at two different excitation energies for two individual QDs of the
small and large QD ensembles, respectively. Upon excitation at 1.53 eV, we find characteristic "spectral line sets" throughout the sample
surface. Neutral exciton (X) and biexciton (XX) features are easily identified by the slope of their integrated intensity dependence with
excitation power ($I_{XX}\sim I_{X}^{2}$) [Fig.~\ref{Fig2}(c)]. Yet, the additional spectral lines observed at both sides of the neutral
exciton and showing a linear behavior with power can correspond to either negatively charged ($X^{-n}$) or positively charged ($X^{+n}$)
excitons. In principle, the residual n-type doping of our sample should favor the capture of additional electrons by QDs. However, at low
temperatures, this effect competes with the trapping of the same electrons by ionized donors. This equilibrium can be disrupted, and the
population of free electrons can be increased, by resonantly pumping the optical transitions related to ionized acceptors (($C^-_{As}$, e),
($C^-_{As}$, $D^+$)) at ~1.49 eV. In such situation, we observe that the low energy peaks in both spectral line sets are enhanced, as shown
in Figs.~\ref{Fig2}(a) and~\ref{Fig2}(b). Peaks labeled A and A' are thus related to radiative recombination of negative trions, $X^{-1}$,
with binding energies $E_{X1^-}^{SQD}=$ 7.5 meV for the SQD, and $E_{X1^-}^{LQD}=$ 3.7 meV for the LQD. An additional peak, not observed
exciting above the GaAs barrier, appears now at a lower energy than the negative trion for the large QD [Fig.~\ref{Fig2}(b)]. Following our
argument, it can be tentatively attributed to the emission of negative doubly charged excitons recombining on their triplet state with
$E_{X2_T^-}^{SQD}=$ 5 meV.~\cite{Alen2005} All the other peaks are partially (peaks B and B') or totally (peaks C and C') quenched upon
resonant excitation on the impurity related optical absorption. In our pumping scheme, this is the expected behavior for neutral excitons
(B and B'), biexcitons (C) and positive trions (C').

The splitting energies just found, 7.5/0.0/-1.0 meV for the $X^{-1}/X/XX$ spectral line set, are typical of as-grown small quantum dots
emitting at this energy.~\cite{Rodt2005} For the large QDs, we have found 5.0/3.7/0.0/-1.0 meV for the $X^{-2}_T/X^{-1}/X/X^{+1}$ set,
which also agree with recent results reported for this kind of large quantum dots (less studied in the recent
literature).~\cite{Savio2006,Cade2006}

\begin{figure}[t]
\includegraphics[width=75 mm]{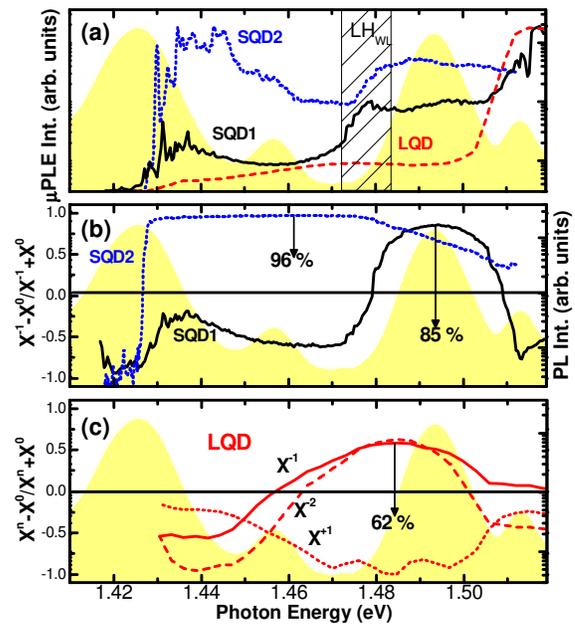}%
\caption{The integrated (see text)  $\mu$PLE spectra obtained at 5 K for two small quantum dots (SQD1-2) and one large quantum dot (LQD)
are shown. The optical pumping efficiency of the different charged exciton complexes is represented for SQD1 and SQD2 in (b) and for LQD in
(c). For comparison, the ensemble PL spectrum shown in Fig. 1(c) has been also included in the background in logarithmic scale (solid
spectrum).} \label{Fig3}
\end{figure}

An enhancement of the negative trion luminescence upon excitation below the GaAs barrier has been reported by Moskalenko \textit{et al},
and by Chang \textit{et al}, for QDs emitting around 1.3 eV.~\cite{Moskalenko2001,Chang2005} To enable the resonant pumping of electrons to
the conduction band, they consider the partial ionization of neutral acceptors by the surface electric field. In our case, electron
transfer towards QDs is warranted by the residual n-type doping of our sample. On one hand, it implies a reservoir of $N_A$ ionized
acceptors for the optical pumping scheme explained above. On the other hand, even in absence of light, $N_D-N_A$ donors still contain an
electron ready for being captured by the QDs if they were sufficiently close to the investigated dot. To illustrate the difference between
both effects, in Fig.~\ref{Fig3} we analyze the PLE spectra corresponding to the spectral line sets identified in Fig.~\ref{Fig2} (LQD and
SQD1), and, for comparison, we also include the PLE spectrum of a different dot (SQD2) which exhibit a clear signature of electron
injection from a donor impurity.

First, in Fig.~\ref{Fig3}(a) we show the integrated  PLE spectra obtained adding up the intensity of all different lines detected for each
dot. All three spectra have been normalized to their maxima and exhibit spectral features clearly correlated with the emission bands shown
in Fig.~\ref{Fig1}(c) and included, as a shadowed spectrum, in Fig.~\ref{Fig1}. Together with the GaAs and heavy hole WL ($HH_{WL}$)
transitions, we found strong absorption at ~1.49 eV and 1.46 eV, and most remarkably at 1.477 eV, which we assign to the light hole WL
transition ($LH_{WL}$) reported at this energy.~\cite{Winzer2002} Yet, the most important conclusions regarding the charge switching effect
can be extracted from panels (b) and (c) of the same figure. We calculate the optical pumping efficiency for the different charged exciton
complexes by evaluating the intensity ratio $\eta=\frac{X^n-X^0}{X^n+X^0}$ as a function of the excitation energy. In Fig.~\ref{Fig3}(b),
we observe that for the negative trion $\eta$ finds a clear maximum at 1.493 eV for SQD1 (solid line), just where the generation of free
electrons is expected through optical pumping of ionized acceptors. The fidelity of the process is 85 \% and spans over a spectral window
of 24 meV (full width at half maximum) around the (e, $C^0_{As}$) band. A similar result is obtained for LQD as shown in Fig. 3(c).
Although, in this case, the effect is less pronounced and occurs in a broader range around 1.485 eV. Out of the impurity window, the
pumping efficiency for $X^{-1}$ decreases and finds its minimum at the GaAs and WL band edges. The behavior of SQD2 is strikingly different
as shown by dashed line in the same figure, and less frequent among the SQDs studied in the sample. With a similar emission energy (1.294
eV) and binding energy (6 meV), the pumping efficiency of the negative trion for SQD2 exhibits an almost flat dependence with excitation
energy. The high fidelity (96 \%) only drops appreciably below the $HH_{WL}$ transition and towards the GaAs barrier. The most likely
explanation for this behaviour is the continuous injection of a spectator electron from a nearby neutral donor with a yield higher than the
radiative rate of the neutral exciton. Our result indicates that in applications that would need the preparation of charged exciton
complexes with high fidelity, modulation doping of the active region can surpass other mechanisms in a broad excitation window, yet the
optical pumping scheme at the acceptor level can be more flexible when more than one complex has to be addressed in the same quantum dot.

Finally, it should be noted that in both SQDs and the LQD, the negative species are largely depleted near the band edges. Indeed, for the
latter we can follow the pumping efficiency of the positive trion to find the opposite trend, as shown by the dotted line in
Fig.~\ref{Fig3}(c). At the band edges, the local density of states is very large and excitons are photocreated with nearly zero momentum.
On our sense, one possible explanation is that a large number of photocreated electrons could be trapped on the ionized donors before
relaxing into quantum dots far away. This would lead to a decreased population of electrons inside the quantum dots. Assuming a shorter
capture time for electrons than for holes in their respective ionized impurities, in average, this process will produce a higher
probability of neutral or positive trion recombination at the band edges. This is a reasonable hypothesis given the larger concentration
and shallower binding energy of donors in our case.

In summary, optical pumping of charged exciton complexes in single InAs QDs has been demonstrated. In the presence of acceptor and donors
in the surroundings of the QDs, exciton charge preparation can be nicely controlled by using photons of the appropriate energy. The charge
mechanism has been compared for two kinds of QDs, and for two different regimes of carrier injection, finding a consistent behavior in
both. We demonstrate that the pumping fidelity can exceed 85 \% enabling the precise control of the charge state in quantum information
applications.

The authors gratefully acknowledge financial support by the Spanish MEC through projects TEC-2005-05781-C03-03 and NAN 2004-09109-C04-04,
by the Italian MIUR through the FIRB project "Nanotecnologie e Nanodispositivi per la Società dell'Informazione" and by the European
Commission through SANDiE Network of Excellence (NMP4-CT-2004-500101).


\end{document}